\documentclass[prb,twocolumn,amsmath,amssymb,superscriptaddress,nofootinbib]{revtex4-2}

\usepackage{graphicx}
\usepackage{amsmath}
\usepackage{yhmath}
\usepackage{bm}
\usepackage{xcolor}
\usepackage{epstopdf}
\usepackage{subfigure}
\usepackage{hyperref}
\usepackage[T1]{fontenc}
\usepackage{lmodern}

\begin{document}
\title{Quantized perfect transmission in graphene nanoribbons with random hollow adsorbates}

\author{Jia-Le Yu}
\affiliation{Hunan Key Laboratory for Super-microstructure and Ultrafast Process, School of Physics, Central South University, Changsha 410083, China}

\author{Zhe Hou}
\affiliation{School of Physics and Technology, Nanjing Normal University, Nanjing 210023, China}

\author{Irfan Hussain Bhat}
\affiliation{Hunan Key Laboratory for Super-microstructure and Ultrafast Process, School of Physics, Central South University, Changsha 410083, China}

\author{Pei-Jia Hu}
\affiliation{Hunan Key Laboratory for Super-microstructure and Ultrafast Process, School of Physics, Central South University, Changsha 410083, China}

\author{Jia-Wen Sun}
\affiliation{Hunan Key Laboratory for Super-microstructure and Ultrafast Process, School of Physics, Central South University, Changsha 410083, China}

\author{Xiao-Feng Chen}
\affiliation{Hunan Key Laboratory for Super-microstructure and Ultrafast Process, School of Physics, Central South University, Changsha 410083, China}
\affiliation{School of Physical Science and Technology, Lanzhou University, Lanzhou 730000, China}

\author{Ai-Min Guo}
\email[]{aimin.guo@csu.edu.cn}
\affiliation{Hunan Key Laboratory for Super-microstructure and Ultrafast Process, School of Physics, Central South University, Changsha 410083, China}

\author{Qing-Feng Sun}
\affiliation{International Center for Quantum Materials, School of Physics, Peking University, Beijing 100871, China}
\affiliation{Hefei National Laboratory, Hefei 230088, China}
\date{\today}

\begin{abstract}
Impurities exist inevitably in two-dimensional materials as they spontaneously adsorb onto the surface during fabrication, usually exerting detrimental effects on electronic transport. Here, we focus on a special type of impurities that preferentially adsorb onto the hollow regions of graphene nanoribbons (GNRs), and study how they affect the quantum transport in GNRs. Contrary to previous knowledge that random adatoms should localize electrons, the so-called Anderson localization, noteworthy quantized conductance peaks (QCPs) are observed at specific electron energies. These QCPs are remarkably robust against variations in system size, GNR edge, and adatom properties, and they can reappear at identical energies following an arithmetic sequence of device width. Further investigation of wavefunction reveals a unique transport mode at each QCP energy which transmits through disordered GNRs reflectionlessly, while all the others become fully Anderson localized, indicating the survival of quantum ballistic transport in the localized regime. Our findings highlight the potential utility of hollow adatoms as a powerful tool to manipulate the conductivity of GNRs, and deepen the understanding of the interplay between impurities and graphene.
\end{abstract}

\maketitle


\textit{Introduction}. Graphene, the first two-dimensional (2D) material initially exfoliated in laboratories \cite{KSNAKGsci,AKGsci,ZSZY,ACFJC,ZSHC,18OGSWXY,18DJRGV,21LCGY,22JYCT,22YZRPC,23SJTN}, has undergone extensive studies during the last two decades in view of its massless Dirac fermions \cite{05YZYW,05KSNAKGnat,07KNAHM,09AHCN,11JHCFC,12TLLLMH,20vAsmC}, which provides an ideal platform for investigating quantum relativistic physics \cite{11JHCFC,12KKGWM,18TJDH,21YNRQC}. With the advent of topology in condensed-matter physics, the discovery of the quantum Hall effect \cite{VPGSGS,YZYW,BOPJH,JRWLD,WLQFS} and the quantum spin Hall effect \cite{CLKEJM,10ZQSAY,10qfSxcX,14JBGKWk} in graphene under strong magnetic fields and graphene on substrate with large spin-orbit coupling, respectively, has endowed graphene with topological property. Moreover, experimental observations have demonstrated that stacking two monolayer graphene sheets and slightly twisting one layer results in the formation of flat bands, where the electron-electron correlation effects dominate. This has led to the emergence of strong-correlation phenomena such as superconducting \cite{18YCVFSF,20yCdCdRL,21YCDRLJMP,21JMPYCKW,21yCjmPkW,22JMPYCLQX} and Mott insulating phases \cite{18HCPLZ,18LRPM}, paving the way for studying ``twistronics'' in 2D materials.

\begin{figure*}
\includegraphics[width=17.2cm]{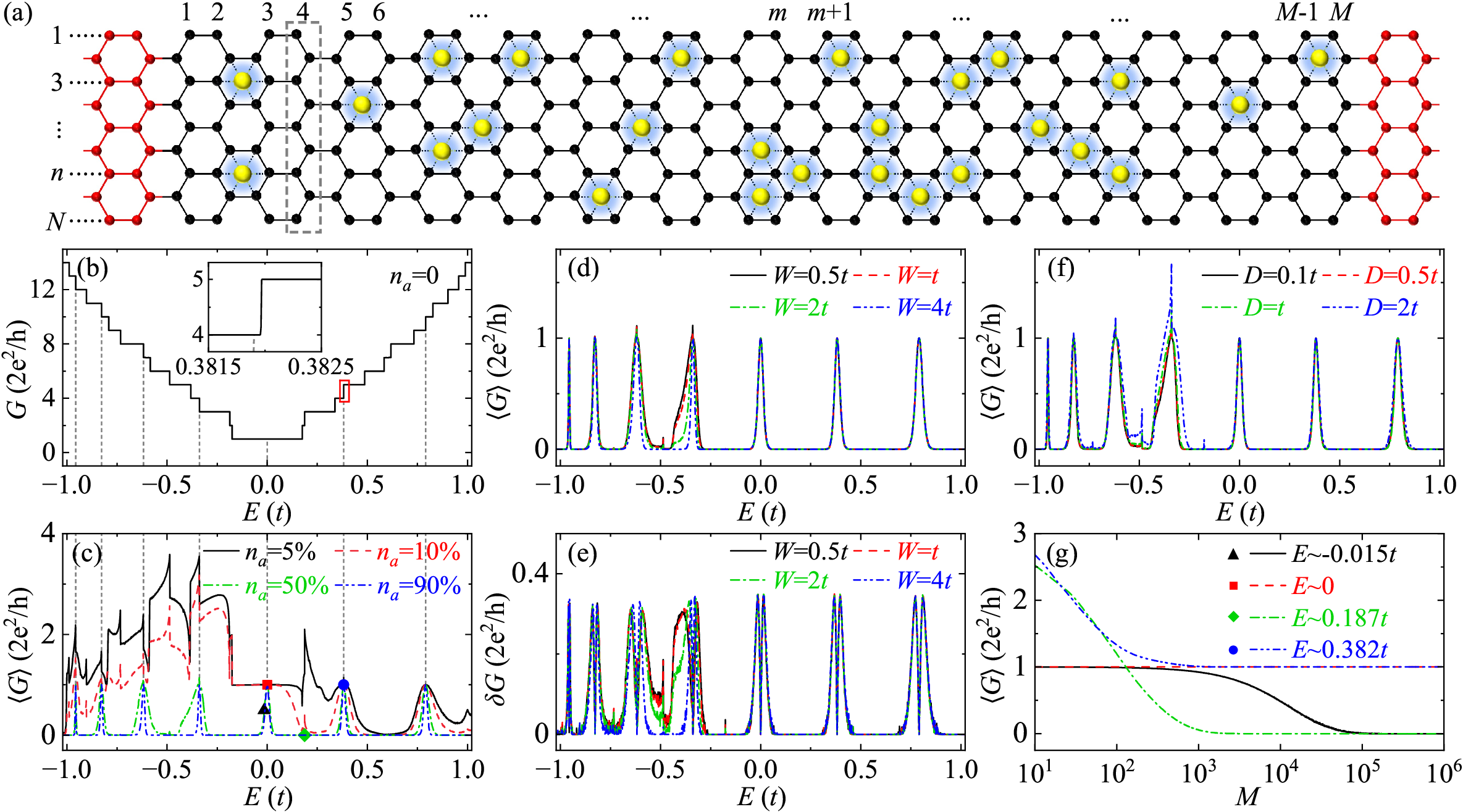}
\caption{\label{fig1} Structure and electron transport property of disordered armchair GNRs. (a) Schematic of a disordered GNR device coupled to left and right semi-infinite GNRs. Here, the black and red balls denote carbon atoms, and the yellow ones are adatoms situated randomly at the hollow regions. The device size is described by the number of slices, $M$, and the number of carbon atoms in each slice, $N$, as indicated by the rectangle. (b) Energy-dependent conductance $G$ of the pristine GNR, with the inset showing the magnified view in the rectangle. Energy-dependent averaged conductance $\langle G \rangle$ of disordered GNRs for (c) several adatom concentrations $n_a$, (d) different on-site energy disorder strength $W$, and (f) different bond disorder strength $D$. (e) Standard deviation $\delta G$ referring to (d). (g) $\langle G \rangle$ versus length $M$ at the energies marked by different symbols in (c). The parameters are $N=29$, $M=10^4$, $n_a=50\%$, $\epsilon_ \alpha=0$, and $\gamma_\alpha=t$, unless specified in the figure.}
\end{figure*}

Graphene's quantum transport in the presence of impurities is also known to be unique \cite{04AHKSAG,13QTZZC,14CHAWZS,16YLYKPL,18RJDE,19KRNDM,19lLjZhS,20FQYZYN,20WDHWHC,20wKsD}. In the presence of long-ranged impurities, graphene maintains high mobility due to Klein tunneling \cite{MIKKN,CWJ,AFYPK,NSBH}, but it experiences Anderson localization with short-ranged random impurities when the inter-valley scattering is strong \cite{08JPRHS,14FGGA,14SGWR,20YGPVML,22YGPVML}. Recently, the study of adatoms onto graphene's surface has garnered much interest, as they can significantly alter graphene's properties by simply manipulating the distribution and type of adatoms \cite{11smKfV,17zLwQjZ,23tTkMnH,10cCmWjJ,11toWaiL,21wHcWhT,21oDlZmY,18kHhMtN}. In general, adatoms adhere to graphene at three different locations \cite{08ktCjbN,11cWjHjA,12hJzhQ,13tEmWmG,18mLcEtL,18sIdKjL,19baBajB,20ygPvmL,22sKtFtP}: top sites (atop carbon atoms), bridge sites (between two adjacent carbon atoms), and hollow sites (at the center of hexagons). Adatoms at top sites have been demonstrated to induce band gaps \cite{10jmGL,10daAavS,19fMndD,20LCFO}, and those at bridge sites are known to induce magnetic moments \cite{03POLASF,09YGZXTZ,19JNIV}. Adatoms occupying hollow sites are usually associated with heavy adatoms  \cite{11cWjHjA,12dMaVmrS,13tEmWmG,16dvTjmM,19iSrY,22fsO} [see the yellow balls in Fig.~\ref{fig1}(a)], which have been shown to induce more intriguing and versatile quantum phases in graphene, such as the spin Hall effect \cite{11cWjHjA,16dvTjmM,18fjSdaB,21sjJmcW}, topological insulator \cite{12JHJA,12HJZHQ,14PHCMSB}, and even superconductivity \cite{07BUAHCN,12GPMC,15BMLGL,16JJZERM}. However, the impact of hollow adatoms on graphene's properties, particularly the quantum transport, has rarely been explored to date.

In this paper, we study theoretically the electron transport through a two-terminal graphene nanoribbon (GNR) with randomly distributed hollow adatoms, as depicted in Fig.~\ref{fig1}(a). Although the overall electron transmission is dramatically reduced, a number of quantized conductance peaks (QCPs) with the value $2e^2/h$ emerge at specific electron energies. These QCPs remain unchanged by varying nanoribbon length and edge, adatom properties, and more importantly they persist in the Anderson localization regime at high adatom concentrations. Furthermore, identical QCPs are observed for various nanoribbon width which follows the arithmetic progression rule. By analysing the wavefunction of carbon atoms in each hexagon, the QCPs can be understood rather simply from the renormalized Shr\"{o}dinger equation. The sum of wavefunction of the six carbon atoms is exactly zero at these QCP energies, thus evanishing the influence of adatoms and making the transmission ballistic. These results unravel, for the first time, the coexistence of two extreme quantum transport phenomena, quantized ballistic transport and Anderson localization, on the same platform, and may facilitate designing novel graphene devices based on impurities instead of their detrimental effects.

\textit{Model}. The electron transport through disordered GNRs, with randomly distributed adatoms at hollow sites, can be described by the tight-binding Hamiltonian:
\begin{equation}
\begin{aligned}
\mathcal{H} = -t\sum_{\langle {i,j} \rangle } {c_i^\dagger } {c_j} + \sum_ \alpha {\epsilon_\alpha }d_\alpha ^\dagger {d_\alpha } + \sum_{\langle {\alpha, j} \rangle } {{\gamma _\alpha }(d_\alpha ^\dagger c_{j } + c_{j}^ \dagger {d_\alpha })}. \label{eq1}	
\end{aligned} \tag{1}
\end{equation}
Here, $c_i^{\dagger}$ ($c_i$) is the creation (annihilation) operator of an electron at site $i$ of graphene lattice, $\left\langle i,j \right\rangle$ denotes all the nearest-neighbor sites with $t$ the hopping integral, and the on-site energy of graphene is taken as the energy reference point. $\epsilon _\alpha$ is the on-site energy of adatom at site $\alpha$ with $d_\alpha ^\dag$ ($d_\alpha$) the creation (annihilation) operator. The last term denotes the coupling between adatom $\alpha$ and all the nearest-neighbor carbon atoms with $\gamma _\alpha$ the isotropic hopping integral which only depends on adatoms.

From the Landauer-B\"{u}ttiker formula \cite{95sD,23pjHjtD}, the conductance of the two-terminal GNR is obtained as $G = (2e^2/h) {\rm Tr}[ {\mathbf \Gamma}_{\rm L} {\mathbf G}^{\rm r} {\mathbf \Gamma }_{\rm R} {\mathbf G}^{\rm a}]$, with ${\mathbf G} ^{\rm r} (E)= [{\mathbf G} ^{\rm a} (E)]^ \dagger=[E{\mathbf I} -{\mathbf H}_ {\rm c}- {\mathbf \Sigma} _{\rm L} ^{\rm r}-{\mathbf \Sigma} _{\rm R} ^{\rm r}]^ {-1}$ the Green's function, and ${{\mathbf{\Gamma }}_{\rm{L/R}}} = i( {{\mathbf{\Sigma }}_{\rm{L/R}}^{\rm{r}} - {\mathbf{\Sigma }}_{\rm{L/R}}^{\rm{a}}} )$ the linewidth function. Here, $E$ is the electron energy, $\mathbf{H}_ {\rm c}$ the Hamiltonian of the central scattering region (CSR), and ${\mathbf \Sigma }_{\rm L/R}^{\rm r(a)}$ the retarded (advanced) self-energy due to the coupling to the left/right semi-infinite GNRs. In the numerical calculations, the hopping integral of graphene is chosen as the energy unit, $t = 1$, the on-site energy of adatoms as $\epsilon_\alpha=0$, and the hopping integral between carbon atoms and adatoms as $\gamma_ \alpha=t$. The CSR size is set to $N=29$ and $M=10^4$, as shown in Fig.~\ref{fig1}(a). Finally, the adatom concentration, defined as the ratio of the number of adatoms to that of hexagons in the CSR, is taken as $n_a= 50 \%$. The numerical results are obtained from $2 \times10^7/M$ disordered samples, and all these parameters will be used throughout the paper, unless stated otherwise.

\textit{The robustness of QCPs.} We first study the electron transport through armchair GNRs by varying the adatom concentration from  $n_a=0$ to $90\%$, as shown in Figs.~\ref{fig1}(b) and~\ref{fig1}(c). As compared with the pristine GNR [Fig.~\ref{fig1}(b)], the introduction of adatoms results in dramatic reduction of the electron transmission along disordered GNRs [Fig.~\ref{fig1}(c)], and there exist zero conductance plateaus at high adatom concentration of $n_a = 50\%$ and $90\%$, a sign of Anderson localization induced by the scattering from randomly distributed adatoms \cite{07SJXYX,08MEIVZ,10MYHJCB,14FGGA,21HYJZ}. Besides, the conductance $\langle G\rangle$ is asymmetric with respect to the line $E=0$ because of the broken electron-hole symmetry induced by the coupling between carbon atoms and hollow adatoms \cite{08JPRHS,15KKBYCW,20YZQG}. Interestingly, a number of QCPs with conductance quantum emerge within the transmission spectrum and locate around the plateau transition points [see the inset of Fig.~\ref{fig1}(b)]. These QCPs exhibit the electron-hole asymmetry as well, where at the electron side the QCPs arise for small $n_a$ and at the hole side they only manifest for large $n_a$. All these phenomena still hold for disordered GNRs with zigzag edge (see the Supplementary Material \cite{SI}), and we will focus on armchair GNRs in the following.

We then investigate how the on-site energy of adatoms and their coupling to the neighboring carbon atoms affect the QCPs. Figures~\ref{fig1}(d) and~\ref{fig1}(f) show the averaged conductance $\langle G\rangle$ versus $E$ by considering the most disordered situation, where the on-site energy $\epsilon_\alpha$ distributes uniformly in $[-W /2, W /2]$ and the coupling parameter $\gamma_\alpha$ within $[t- D/2, t+ D/2]$, with $W$ and $D$ the on-site energy and bond disorder strengths, respectively. This refers to real situations that the adatoms can vary in both size and vertical distance from graphene plane \cite{14vSsS,21fIcF,23jpCjB}. We can see that the profile of most QCPs remains the same by changing $W$ or $D$ [Figs.~\ref{fig1}(d) and~\ref{fig1}(f)]. By contrast, the two QPCs at $E \sim -0.618t$ and $-0.338t$ seem to be sensitive to the adatoms, where $\langle G\rangle$ decreases with increasing $W$ and increases with $D$. The latter anomalous behavior originates from the fact that, with increasing $D$, the adatoms will effectively decouple from the GNR and the adatom concentration is declined, leading to the increment of the transmission ability, because of the finite-size effect. Indeed, all these QCPs remain unchanged for sufficiently long GNRs, no matter the values of $W$ and $D$, implying the robustness against the adatoms.

Figure~\ref{fig1}(e) plots the energy-dependent standard deviation $\delta G \equiv \sqrt { \langle G^2 \rangle - {\langle G \rangle }^2}$, in accordance with Fig.~\ref{fig1}(d). There always exist dips of $\delta G=0$ at the QCP positions and two deviation peaks of $\delta G \sim 0.66e^2/h$ for each QCP, indicating the QCPs can be observed in any disordered GNR sample. When the electron energy is far away from the QCP positions, the standard deviation satisfies $\delta G=0$ [see the electron side in Fig.~\ref{fig1}(e)], indicating the localization behavior in any disordered GNR with length $M=10^4$. Figure~\ref{fig1}(g) displays $\langle G \rangle$ versus $M$ for typical electron energies marked by different symbols in Fig.~\ref{fig1}(c). It is clear that, at the QCP positions, $\langle G \rangle$ usually decreases with increasing $M$ and then saturates at conductance quantum in the large length limit [see the red-dashed and blue-dash-dot-dotted lines in Fig.~\ref{fig1}(g)], implying the quantum ballistic transport behavior of the QCPs. When $E$ deviates from the QCP positions, $\langle G \rangle$ gradually decreases with increasing $M$ and finally becomes zero, which corresponds to the Anderson localization for non-QCP energies.

\begin{figure}
\includegraphics[width=8.6cm]{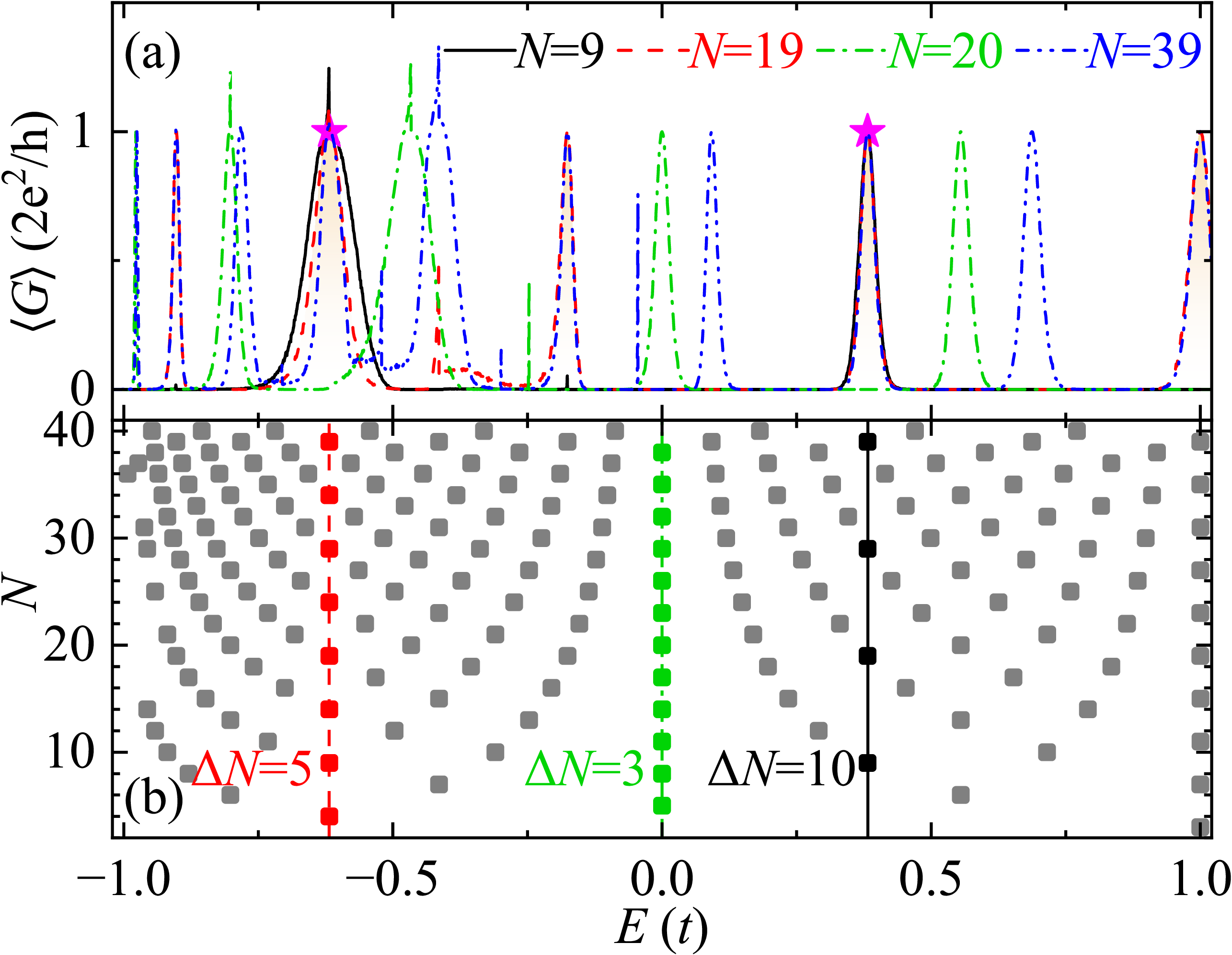}
\caption{\label{fig2} Electron transport along disordered armchair GNRs with different width. (a) Energy-dependent $\langle G\rangle$ for different width $N$. (b) Evolution of the QCPs for different $N$. Here, $n_a = 50\%$ and the other parameters are the same as Fig.~\ref{fig1}(c).}
\end{figure}

\textit{Width-dependent QCPs}. We then study the electron transport through disordered GNRs by considering the nanoribbon width, as shown in Fig.~\ref{fig2}(a), where $\langle G \rangle$ versus $E$ is displayed for typical values of $N$. One can see that the number of QCPs increases with $N$, owing to the increment of transport modes. Some QCPs for $N=39$ overlap all those for $N=19$, while the remaining QCPs overlap all those for $N=20$ by properly shifting their positions. Interestingly, the two QCPs locate at the same energies for $N=9$, 19, and 39 [see the stars in Fig.~\ref{fig2}(a)]. This phenomenon can also be detected in other disordered GNRs with various width, as shown in Fig.~\ref{fig2}(b), where the evolution of all the QCPs in the energy region $[-t, t]$ is displayed by ranging the width from $N=3$ to 40. For example, identical QCPs are observed at $E \sim -0.618t$ for $N= 4,9,14,...$, at $E \sim 0$ for $N= 5, 8, 11,...$, and at $E \sim 0.382t$ for $N= 9, 19, 29,...$ [see the red-dashed, green-dash-dotted, and black-solid lines in Fig.~\ref{fig2}(b)]. This evolution phenomenon can be formulated in an arithmetic sequence of $N_j = j \Delta N -1$, with $j$ an integer and $\Delta N$ the width difference between two disordered GNRs of identical QCPs.

\begin{figure}
\includegraphics[width=8.6cm]{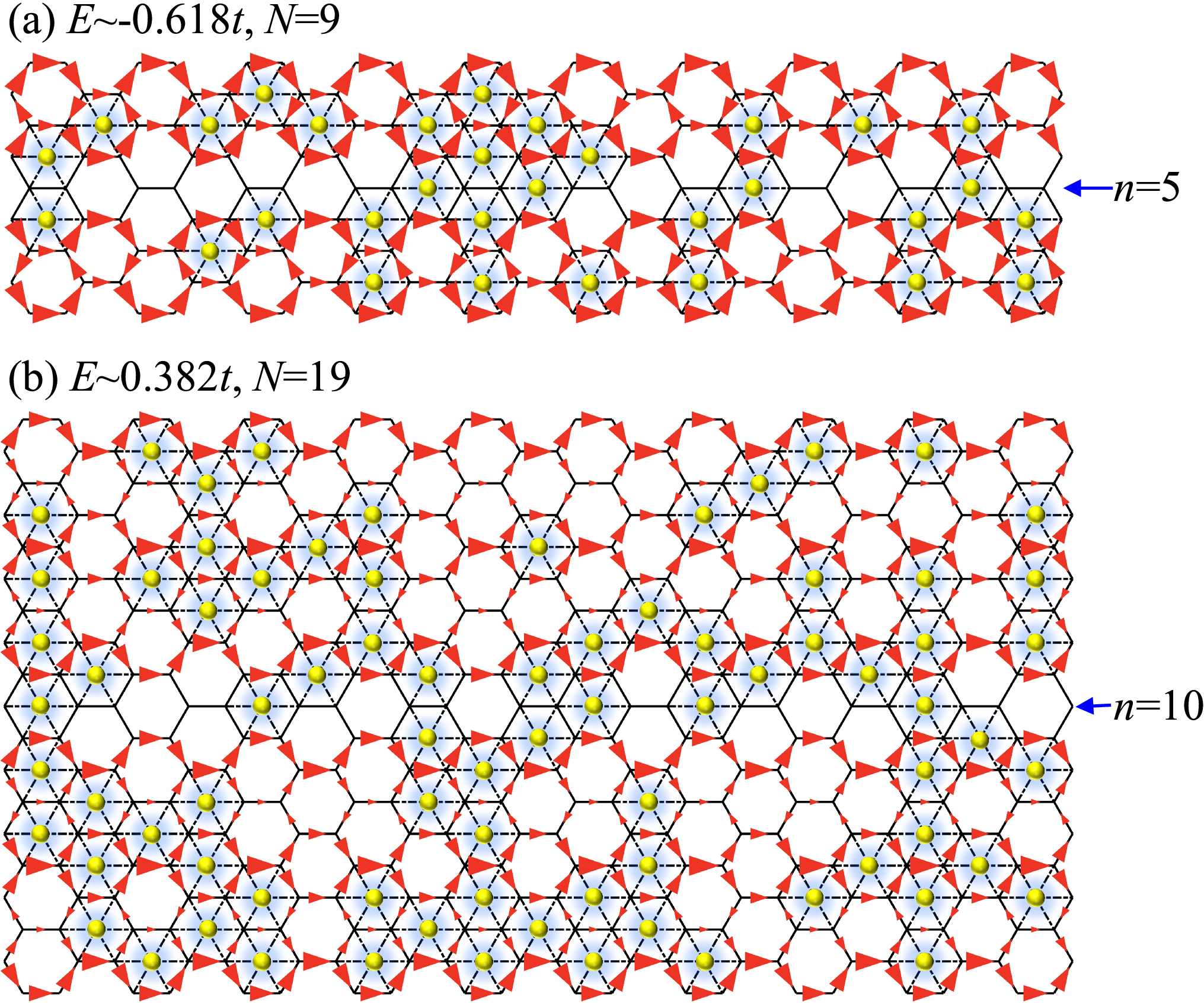}
\caption{\label{fig3} Spatial distributions of bond currents for the QCPs at (a) $E \sim -0.618t$ with $N=9$ and (b) $E \sim 0.382t$ with $N=19$, as marked by the stars in Fig.~\ref{fig2}(a). The two panels refer to the middle segment of disordered GNRs with $M = 2 \times 10^4 + 20$, and the arrow size is proportional to the magnitude of bond currents.}
\end{figure}

To understand the width-dependent QCPs, Figs.~\ref{fig3}(a) and~\ref{fig3}(b) show the distributions of bond currents for disordered GNRs at the two QCP positions marked by the stars in Fig.~\ref{fig2}(a), which can be calculated from the lesser Green's function \cite{09hJlwqfS,13chLerM,22pjHtfF,eq_current}. Here, the arrow size is proportional to the magnitude of bond currents. One can see that the currents only flow along the bonds between neighboring carbon atoms, and thus the scattering from the adatoms disappears, leading to the emergence of the QCPs. Besides, one can identify other important features from Fig.~\ref{fig3}. (i) The spatial mirror symmetry is maintained with respect to the $(N+1)/2$-row for odd $N$. (ii) The fifth row for $N=9$ and the tenth row for $N=19$, where the bond currents are zero, can be used to divide the disordered GNR into two isolated segments with identical distributions of bond currents. As a result, identical QCPs can be observed at $E \sim -0.618t$ for $N_j=4$ and $9$, and at $E \sim 0.382t$ for $N_j =9$ and $19$. We conclude that, by applying the zero-current row to divide the disordered GNR into the smallest segment, the width at which the quantized perfect transmission happens with identical QCPs can be determined.

\begin{figure}
\includegraphics[width=8.6cm]{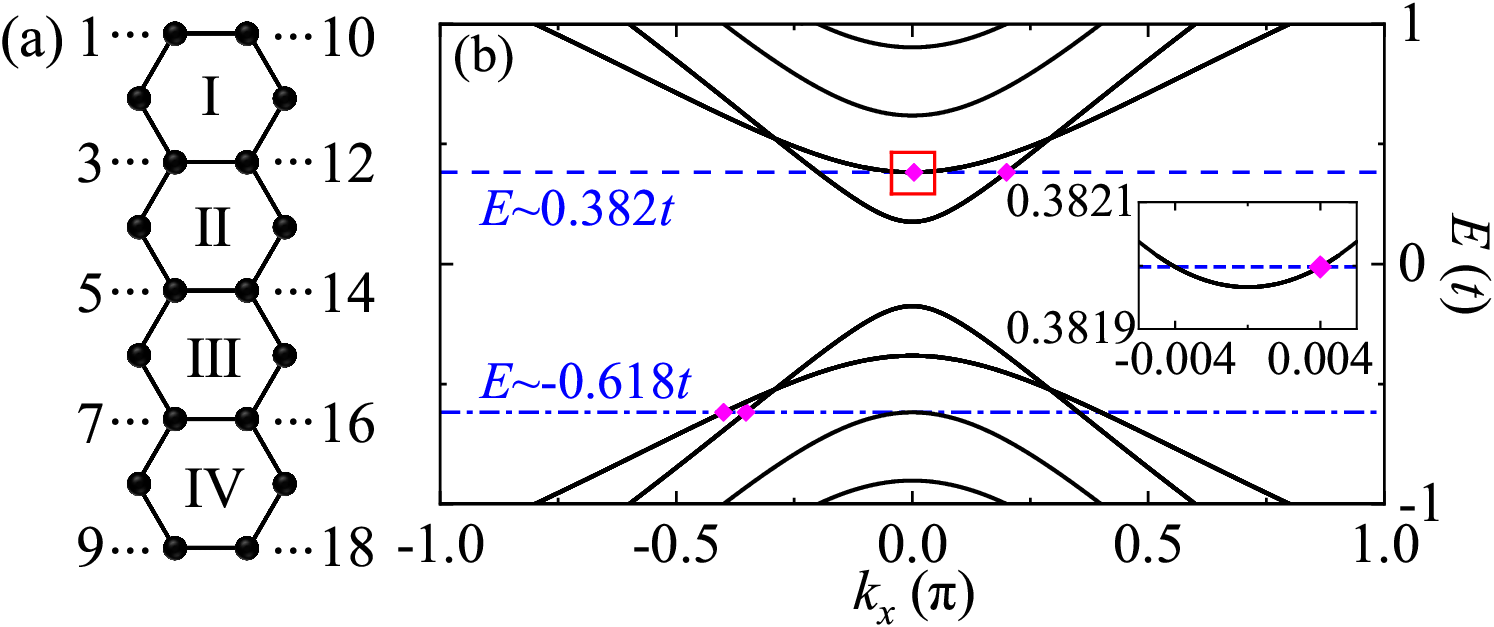}
\caption{\label{fig4} (a) Unit cell and (b) dispersion relation of armchair GNR with $N=9$. The carbon atoms and hexagons in (a) are described by the Arabic and Roman numerals, respectively. The magenta diamonds in (b) refer to specific momenta at which energies the QCPs take place, as marked by the stars in Fig.~\ref{fig2}(a).}
\end{figure}

\begin{table}
\caption{\label{table1} Sum of wavefunction of the six carbon atoms in every hexagon at different momenta $k_x$, as marked by the magenta diamonds in Fig.~\ref{fig4}(b).}
\begin{ruledtabular}
\begin{tabular}{l r r r r}
& \multicolumn{2}{c}{$E\sim-0.618t$}           & \multicolumn{2}{c}{$E\sim0.382t$}    \\ \cline{2-3}\cline{4-5}
& \multicolumn{1}{c}{$k_x=-0.4\pi$} & \multicolumn{1}{c}{$k_x=-0.352\pi$} & \multicolumn{1}{c}{$k_x=0.2\pi$} & \multicolumn{1}{c}{$k_x=0.004\pi$} \\  \hline
\uppercase\expandafter{\romannumeral1}  & \multicolumn{1}{c}{0}    & $0.255-0.256i$     & \multicolumn{1}{c}{0}     & $0.000-0.003i$  \\
\uppercase\expandafter{\romannumeral2}  & \multicolumn{1}{c}{0}    & $-0.158+0.158i$    & \multicolumn{1}{c}{0}     & $0.000+0.005i$  \\
\uppercase\expandafter{\romannumeral3}  & \multicolumn{1}{c}{0}    & $-0.158+0.158i$    & \multicolumn{1}{c}{0}     & $0.000-0.005i$  \\
\uppercase\expandafter{\romannumeral4}  & \multicolumn{1}{c}{0}    & $0.255-0.256i$     & \multicolumn{1}{c}{0}     & $0.000+0.003i$  \\
\end{tabular}
\end{ruledtabular}
\end{table}

\textit{The origin of QCPs}. To further elucidate the physical origin of the QCPs, we calculate the wavefunction of specific electronic states at which the QCPs take place. Figures~\ref{fig4}(a) and~\ref{fig4}(b) plot, respectively, the unit cell and the dispersion relation of the pristine GNR with $N = 9$. Here, the carbon atoms and hexagons are labeled by Arabic and Roman numerals, respectively. The blue dash-dotted and dashed lines correspond to the QCP positions [see the stars in Fig.~\ref{fig2}(a)], and these two lines independently intersect the dispersion relation, as shown by the magenta diamonds and the inset in Fig.~\ref{fig4}(b). Table~\ref{table1} displays the sum of wavefunction of the six carbon atoms in every hexagon at different momenta $k_x$, where the wavefunction of all the carbon atoms is shown in the Supplementary Material \cite{SI}. We can see from Tab.~\ref{table1} that, at the QCP positions, there always exists one specific momentum $k_x$, at which the sum of wavefunction is zero exactly.

We then write down the Schr\"{o}dinger equation on the carbon atoms:
\begin{equation}
\begin{aligned}
E{\psi _i} =  - t\sum\limits_{{\langle j \rangle }_i} {\psi _j}  + \sum \limits_{{{\langle \alpha \rangle }_i}} {\gamma _\alpha }{\varphi _\alpha } , \label{eq2}
\end{aligned} \tag{2}
\end{equation}
where $\psi _i$ is the wavefunction of carbon atoms at position {\bm{$r}_i$}, $j$ is the site index for all the neighboring carbon atoms around site $i$, and $\alpha$ is the one for all the neighboring adatoms to $i$, with $\varphi _\alpha$ the wavefunction of adatoms. Similarly, for the adatoms, we have $(E-\epsilon_ \alpha) \varphi_\alpha =\gamma_ \alpha\sum\nolimits_{{{\langle {j'} \rangle }_\alpha }} {{\psi _{j'} }}$. Substituting $\varphi _\alpha$ into Eq.~(\ref{eq2}), we derive
\begin{equation}
	\begin{aligned}
		E{\psi _i} =  - t\sum\limits_{{{\left\langle j \right\rangle }_i}} {{\psi _{j}}}  + \sum\limits_{{{\left\langle \alpha \right\rangle }_i}}  {\frac{{\gamma _\alpha ^2}}{{E - {\epsilon _\alpha }}}\sum\limits_{{{\left\langle {j'} \right\rangle }_\alpha}} {{\psi _{{j' }}}} }.   \label{eq3}
	\end{aligned} \tag{3}
\end{equation}
It can be deduced from Eq.~(\ref{eq3}) that if the sum of wavefunction $\psi_{j'}$ of the six carbon atoms around the adatom $\alpha$ equals zero, the second term on the right side vanishes. This elucidates why the adatoms have no impact on the electron transport for this special transport mode, where the electrons propagate ballistically through disordered GNRs, leading to a single quantized conductance of $2e^ 2/h$. While for other transport modes, the presence of random adatoms significantly alters their behavior, causing the electrons to become Anderson localized. Consequently, a QCP of conductance quantum emerges. This phenomenon can find its analogy in disordered quantum Hall systems where only the dissipationless Hall edge states contribute to a quantized conductance \cite{95BH,06LBHAF,12FDJBO,21SKJS}, but the physical origin is different.

\textit{Conclusion}. In summary, we investigate the electron transport through GNRs with random hollow adatoms. We uncover unexpected QCPs at specific energies within the transmission spectrum, alongside the overall suppression of conductance. These QCPs are found to be very robust against system size, GNR edge, and adatom properties, and most importantly, they can survive in the presence of Anderson localization. A systematic analysis on distributions of bond currents and the wavefunction reveals the ballistic transport feature of these exotic QCPs. Our findings contribute significantly to the understanding of the interplay between graphene and impurities, offering valuable insights for the design of conductance switches in graphene-based materials utilizing hollow adatoms.

\textit{Ackowledgement}. This work was supported by the National Natural Science Foundation of China (Grant Nos. 12274466, 11874428, 12304070, 12374034, and  11921005), the Hunan Provincial Science Fund for Distinguished Young Scholars (Grant No. 2023JJ10058), the Innovation Program for Quantum Science and Technology
(2021ZD0302403), and the High Performance Computing Center of Central South University.


\begin{references}
\bibitem{eq_current} The local current flowing from the $m$th slice to the $(m + 1)$th one can be written as: ${\bf I}_m=(-2e /h) \int [{{\bf H}_{m,m+1} {\bf G}_{m+1,m}^ < (\xi)- {\bf H}_ {m+1,m} {\bf G}_{m,m + 1}^ < (\xi)} ]d\xi $. Here, ${\bf H}_{m+1,m}$ is the coupling matrice between the $m$th slice and the $(m+1)$th one, and ${\bf G}_{m,m + 1}^ <$ is the interchain lesser Green's function.
\end{references}
\end{document}


\title{Supplementary Information for ``Quantized perfect transmission in graphene nanoribbons with random hollow adsorbates''}

\author{Jia-Le Yu}
\affiliation{Hunan Key Laboratory for Super-microstructure and Ultrafast Process, School of Physics, Central South University, Changsha 410083, China}

\author{Zhe Hou}
\affiliation{School of Physics and Technology, Nanjing Normal University, Nanjing 210023, China}

\author{Irfan Hussain Bhat}
\affiliation{Hunan Key Laboratory for Super-microstructure and Ultrafast Process, School of Physics, Central South University, Changsha 410083, China}

\author{Pei-Jia Hu}
\affiliation{Hunan Key Laboratory for Super-microstructure and Ultrafast Process, School of Physics, Central South University, Changsha 410083, China}

\author{Jia-Wen Sun}
\affiliation{Hunan Key Laboratory for Super-microstructure and Ultrafast Process, School of Physics, Central South University, Changsha 410083, China}

\author{Xiao-Feng Chen}
\affiliation{Hunan Key Laboratory for Super-microstructure and Ultrafast Process, School of Physics, Central South University, Changsha 410083, China}
\affiliation{School of Physical Science and Technology, Lanzhou University, Lanzhou 730000, China}

\author{Ai-Min Guo}
\affiliation{Hunan Key Laboratory for Super-microstructure and Ultrafast Process, School of Physics, Central South University, Changsha 410083, China}

\author{Qing-Feng Sun}
\affiliation{International Center for Quantum Materials, School of Physics, Peking University, Beijing 100871, China}
\affiliation{Hefei National Laboratory, Hefei 230088, China}
\date{\today}

\maketitle
\subsection{1. Electron transport along disordered zigzag graphene nanoribbons with random hollow adsorbates}

\begin{figure}[hbt!]
\includegraphics[width=7.8cm]{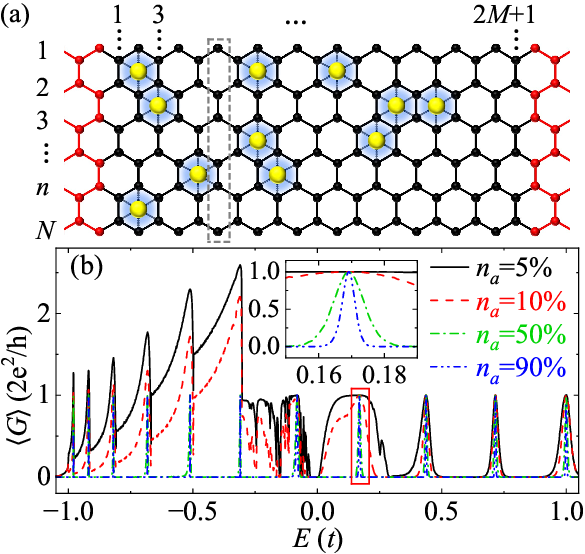}
\caption{\label{figs1} (a) Schematic of a two-terminal zigzag GNR coupled to left and right semi-infinite electrodes. Here, the black and red balls denote carbon atoms, and the yellow ones are adatoms randomly distributed at the hollow sites. (b) Energy-dependent averaged conductance $\langle G\rangle$ for disordered zigzag GNRs with the adatom concentration ranging from $n_{a}=5\%$ to $90\%$. The parameters are $N=14$, $M = 10^4$, $W = 0$, and $D = 0$.}
\end{figure}

Figure~\ref{figs1}(a) displays the schematic of a two-terminal zigzag graphene nanoribbon (GNR) device, while Fig.~\ref{figs1}(b) shows the averaged conductance $\left\langle G \right\rangle$ for disordered zigzag GNRs with several values of the adatom concentrations, as a function of the electron energy $E$. It is clear that the transmission ability is dramatically declined over the energy spectrum and the conductance approaches zero, especially at high adatom concentrations, leading to Anderson localization in disordered zigzag GNRs. Nevertheless, a number of quantized conductance peaks (QCPs) with conductance quantum remain in the energy spectrum. These phenomena are similar as the armchair GNRs in the main text, demonstrating that the QCPs are robust against the GNR edge.

\newpage

\subsection{2. Wavefunction of all the carbon atoms at the QCPs}

\begin{table}[hbt!]
\caption{\label{table2} Wavefunction of all the carbon atoms for specific $k_x$ marked by the magenta diamonds in Fig.~4(b) of the main text, and the sum of wavefunction of the six carbon atoms arranged in the hexagons shown in Fig.~4(a).}
\begin{ruledtabular}
\begin{tabular}{l r r r r}
& \multicolumn{2}{c}{$E\sim-0.618t$}           & \multicolumn{2}{c}{$E\sim0.382t$}    \\ \cline{2-3}\cline{4-5}
& \multicolumn{1}{c}{$k_x=-0.4\pi$} & \multicolumn{1}{c}{$k_x=-0.352\pi$} & \multicolumn{1}{c}{$k_x=0.2\pi$} & \multicolumn{1}{c}{$k_x=0.004\pi$} \\  \hline
				1  &$0.243 - 0.177i$    &$0.000 - 0.256i  $       &$0.079 - 0.243i  $     &$0.301 + 0.003i   $      \\
				2  &$-0.150 - 0.109i$   &$-0.256 - 0.158i $       &$-0.286 + 0.093i $     &$0.186 - 0.001i   $      \\
				3  &$-0.150 + 0.109i$   &$0.000 - 0.098i  $       &$0.030 - 0.093i  $     &$-0.186 - 0.002i  $      \\
				4  &$0.243 + 0.177i$    &$0.158 + 0.098i  $       &$0.177 - 0.057i  $     &$-0.301 + 0.002i  $      \\
				5  &$0.000 + 0.000i$    &$0.000 + 0.316i  $       &$-0.098 + 0.301i $     &$0.000 - 0.000i   $      \\
				6  &$-0.243 - 0.177i $  &$0.158 + 0.098i  $       &$0.177 - 0.057i  $     &$0.301 - 0.002i   $      \\
				7  &$0.150 - 0.109i  $  &$0.000 - 0.098i  $       &$0.030 - 0.093i  $     &$0.186 + 0.002i   $      \\
				8  &$0.150 + 0.109i  $  &$-0.256 - 0.158i $       &$-0.286 + 0.093i $     &$-0.186 + 0.001i  $      \\
				9  &$-0.243 + 0.177i $  &$0.000 - 0.256i  $       &$0.079 - 0.243i  $     &$-0.301 - 0.003i  $      \\
				10 &$0.301 + 0.000i  $  &$0.256 + 0.000i  $       &$0.256 - 0.000i  $     &$-0.301 - 0.000i  $      \\
				11 &$-0.057 + 0.177i $  &$0.158 + 0.256i  $       &$-0.177 + 0.243i $     &$-0.186 - 0.003i  $      \\
				12 &$-0.186 + 0.000i $  &$0.098 + 0.000i  $       &$0.098 - 0.000i  $     &$0.186 - 0.000i   $      \\
				13 &$0.093 - 0.286i  $  &$-0.098 - 0.158i $       &$0.109 - 0.150i  $     &$0.301 + 0.005i   $      \\
				14 &$0.000 + 0.000i  $  &$-0.316 + 0.000i $       &$-0.316 - 0.000i $     &$0.000 - 0.000i   $      \\
				15 &$-0.093 + 0.286i $  &$-0.098 - 0.158i $       &$0.109 - 0.150i  $     &$-0.301 - 0.005i  $      \\
				16 &$0.186 + 0.000i  $  &$0.098 + 0.000i  $       &$0.098 - 0.000i  $     &$-0.186 - 0.000i  $      \\
				17 &$0.057 - 0.177i  $  &$0.158 + 0.256i  $       &$-0.177 + 0.243i $     &$0.186 + 0.003i   $      \\
				18 &$-0.301 + 0.000i $  &$0.256 + 0.000i  $       &$0.256 - 0.000i  $     &$0.301 - 0.000i   $      \\
				\uppercase\expandafter{\romannumeral1}  & \multicolumn{1}{c}{0}    & $0.255-0.256i$     & \multicolumn{1}{c}{0}     & $0.000-0.003i$  \\
				\uppercase\expandafter{\romannumeral2}  & \multicolumn{1}{c}{0}    & $-0.158+0.158i$    & \multicolumn{1}{c}{0}     & $0.000+0.005i$  \\
				\uppercase\expandafter{\romannumeral3}  & \multicolumn{1}{c}{0}    & $-0.158+0.158i$    & \multicolumn{1}{c}{0}     & $0.000-0.005i$  \\
				\uppercase\expandafter{\romannumeral4}  & \multicolumn{1}{c}{0}    & $0.255-0.256i$     & \multicolumn{1}{c}{0}     & $0.000+0.003i$  \\
\end{tabular}
\end{ruledtabular}
\end{table}